\documentclass{nature}
\usepackage{hyperref}
\usepackage{cellspace}
\usepackage[flushleft]{threeparttable}
\usepackage{graphicx}% Include figure files
\usepackage{dcolumn}% Align table columns on decimal point
\usepackage{amsmath}
\usepackage{bm}% bold math
\title{Link prediction via linear optimization}

\author{Ratha Pech$^1$, Dong Hao$^{1,2}$, Yan-Li Lee$^{1,2}$, Ye Yuan$^{3}$ \& Tao Zhou$^{1,2}$}

\begin{document}

\maketitle

\begin{affiliations}
 \item CompleX Lab, University of Electronic Science and Technology of China, Chengdu 611731, People's Republic of China.
 \item Big Data Research Center, University of Electronic Science and Technology of China, Chengdu 611731, People's Republic of China.
 \item School of Automation, the State Key Laboratory of Digital Manufacturing Equipment and Technology, Huazhong University of Science and Technology, Wuhan 430074, China
\end{affiliations}

\begin{abstract}
Link prediction is an elemental challenge in network science, which has already found applications in guiding laboratorial experiments, digging out drug targets, recommending online friends, probing network evolution mechanisms, and so on. With a simple assumption that the likelihood of the existence of a link between two nodes can be unfolded by a linear summation of neighboring nodes' contributions, we obtain the analytical solution of the optimal likelihood matrix, which shows remarkably better performance in predicting missing links than the state-of-the-art algorithms for not only simple networks, but also weighted and directed networks. To our surprise, even some degenerated local similarity indices from the solution outperform well-known local indices, which largely refines our knowledge, for example, the number of 3-hop paths between two nodes more accurately predicts missing links than the number of 2-hop paths (i.e., the number of common neighbors), while in previous methods, longer paths are always considered to be less important than shorter paths.
\end{abstract}

Thanks to the breakthrough in uncovering the structural complexity (e.g., small-world \cite{Watts1998} and scale-free \cite{Barabasi1999} properties) in real networks, the recent twenty years have witnessed an explosion in the studies of networks, which is turning the so-called \emph{network science} from niche branches of science in mathematics (i.e., graph theory) and social science (i.e., social network analysis) to an interdisciplinary focus that attracts increasing attentions from physicists, mathematicians, social scientists, computer scientists, biologists, and so on. Recently, the research focus of network science has been shifting from macroscopic statistical regularities \cite{Newman2010} to different roles played by microscopic elements, such as nodes \cite{Lu2016} and links \cite{Csermely2006}, in network structure and functions. Therein, link prediction is an elemental challenge that aims at estimating the likelihood that a nonobserved link exists, on the basis of observed links in a network \cite{Lu2011}.

Link prediction is of particular significance. Theoretically speaking, link prediction can be used as a probe to quantify to which extent the network formation and evolution can be explained by a mechanism model, since a better model should be in principle transferred to a more accurate algorithm \cite{Wang2012,Zhang2015}. Beyond theoretical interests, link prediction has already found many applications. For example, our knowledge of biological interactions is highly limited, with approximately 99.7\% of the molecular interactions in human beings still unknown \cite{Stumpf2008}. Instead of blindly checking all possible interactions, to predict based on known
interactions and focus on those links most likely to exist can sharply reduce the experimental costs if the predictions are accurate enough \cite{Barzel2013}. Analogously, the known interactions between drugs and target proteins are very limited, while it is believed that any single drug can interact with multiple targets \cite{Chong2007}. By this time, link prediction algorithms have already played a critical role in finding out new uses of old drugs \cite{Ding2013}. Besides dealing with missing data problems, link prediction algorithms can also be used to predict the links that may appear in the future of evolving networks, with obviously commercial values in friend recommendations of online social networks \cite{Aiello2012} and product recommendations in e-commercial web sites \cite{Lu2012}.

Many algorithms have been proposed to solve the link prediction problem, including \emph{probabilistic models} \cite{Neville2007,Yu2007} that establish a model with usually a large number of parameters to best fit the observed data and then predict missing links by using the learned model, \emph{similarity-based algorithms} \cite{Liben-Nowell2007,Zhou2009} that assign a similarity score to every pair of nodes and rank all non-observed links according to their scores, \emph{maximum likelihood methods} \cite{Clauset2008,Guimera2009} that presuppose some network organizing principles with detailed rules and specific parameters being obtained by maximizing the likelihood of the observed structure and then calculate the likelihood of any nonobserved link according to those rules and parameters, and some others \cite{Backstrom2011,Pech2017}. Despite these achievements, how to design effective and efficient algorithms remains a conspicuous challenge. The similarity-based algorithms are often very efficient for its low computational complexity (especially for local similarity indices \cite{Zhou2009}) but less accurate. The maximum likelihood methods are highly time consuming, with typical ones (e.g., hierarchical structure model \cite{Clauset2008}, stochastic block model \cite{Guimera2009} and LOOP model \cite{Pan2016}) can only handle networks with a few thousands of nodes, while real social networks scale from millions to more than a billion nodes. The probabilistic models often require the information about node attributes in addition to the observed network structure, which highly limits their applications. And the number of parameters are too many so that we cannot easily find any insights about network organization.

In this paper, we assume that the likelihood of the existence of a nonobserved link from node $i$ to node $j$ can be unfolded by a linear summation of contributions from $i$'s neighbors. Accordingly, we transfer link prediction to an optimization problem for the likelihood matrix, which can be solved analytically. We have tested our algorithms as well as the state-of-the-art benchmarks in 24 real networks from disparate fields, including 8 simple networks, 8 weighted networks and 8 directed networks. Extensive empirical comparison shows that our algorithms remarkably outperforms the similarity-based algorithm and slightly better than the maximum likelihood methods. At the same time, the time complexity of our algorithm is much lower than the maximum likelihood methods. We further analyze some degenerated local similarity indices for simple networks from the analytical solution, which still perform much better than many well-known local indices. Of particular interest, the direct count of 3-hop paths between two nodes $i$ and $j$, say $(\mathbf{A}^3)_{ij}$ where $\mathbf{A}$ is the adjacency matrix, give more accurate predictions for missing links than the widely used common neighbor index $(\mathbf{A}^2)_{ij}$. This finding shakes a common belief in graph mining that the statistics on shorter paths are more significant than those on longer paths, as indicated by the decaying factor in Katz index \cite{Katz1953} and local path index \cite{Lu2009}.

\section*{Algorithm}

Considering an observed network $G(V,E)$ with $V$ and $E$ being the sets of nodes and links, respectively. The corresponding adjacency matrix $\mathbf{A}$ is defined as $\mathbf{a}_{ij}=1$ if there is a link from node $i$ to node $j$, and $\mathbf{a}_{ij}=0$ otherwise. For simple networks (i.e., undirected unweighted networks), $\mathbf{A}$ is symmetric, say $\mathbf{a}_{ij}=\mathbf{a}_{ji}$; for directed networks, in general, $\mathbf{a}_{ij}$ can be different from $\mathbf{a}_{ji}$; for weighted networks, $\mathbf{a}_{ij}$ denotes the weight assigned to the link from $i$ to $j$, which is not necessarily equal to 1. In the following deviation, we use the general definition of $\mathbf{A}$ so that the results can be directly applied for directed and weighted networks.

We assume that the likelihood of the existence of a link from $i$ to $j$, denoted by $\mathbf{s}_{ij}$, can be unfolded by a linear summation of contributions from $i$'s neighbors, namely
\begin{equation}
\mathbf{s}_{ij}=\sum_k \mathbf{a}_{ik}\mathbf{z}_{kj},
\end{equation}
where $\mathbf{z}_{kj}$ is the contribution from node $k$ to node $j$. In the likelihood matrix $\mathbf{S}$ ($\mathbf{S=AZ}$, as defined in Eq. (1), which is also named as score matrix or similarity matrix in similarity-based algorithms), only the elements corresponding to nonobserved links are meaningful in link prediction, but the elements corresponding to observed links can be used to evaluate the rationality of $\mathbf{S}$, because to be self-consistent, if $\mathbf{a}_{ij}>\mathbf{a}_{pq}$, $\mathbf{s}_{ij}$ should also be larger than $\mathbf{s}_{pq}$. That is to say, the difference between $\mathbf{A}$ and $\mathbf{S}$ should be small. At the same time, to avoid overfitting, the magnitude of $\mathbf{Z}$ should also be small. Accordingly, the determination of the likelihood matrix $\mathbf{S}$ can be simply transferred to an optimization problem
\begin{equation}
   \min_\mathbf{Z} \alpha||\mathbf{A-AZ}|| + ||\mathbf{Z}||,
\end{equation}
where $\alpha$ is a free parameter that balances the two requirements and $||\cdot||$ denotes a certain matrix norm.

\begin{figure}
    \centering
	\includegraphics[width=1\textwidth]{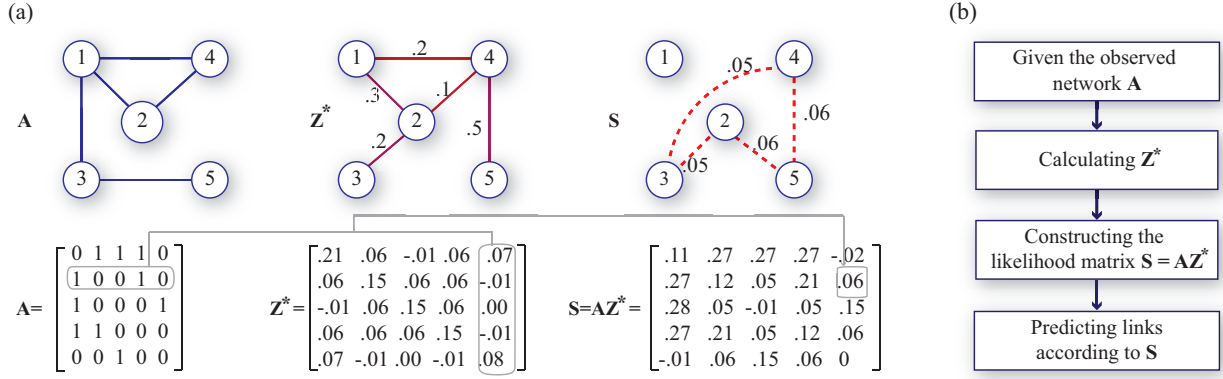}
    \caption{(a) The illustration of how to calculate $\mathbf{Z}^*$ and $\mathbf{S}$ for a small-size simple network. According to the proposed algorithm, the two links $(2,5)$ and $(4,5)$ are considered to be most likely missing links. (b) The whole procedure of the algorithm. }
\label{fig:toy}
\end{figure}

To make Eq. (2) solvable, we choose the Frobenius norm with power 2, namely to minimize
\begin{equation}
\mathbf{E}=\alpha||\mathbf{A-AZ}||_F^2 + ||\mathbf{Z}||_F^2,
\end{equation}
where $||\mathbf{X}||_F^2=\mathbf{Tr}(\mathbf{X}^T\mathbf{X})$. Performances of other commonly used norms, such as $\ell_1$-norm and nuclear norm are similar to the above one (see Supplementary Note 1). The expansion of Eq. (3) reads
\begin{equation}
\begin{split}
\mathbf{E} &=  \alpha \mathbf{Tr}[(\mathbf{A-AZ})^T(\mathbf{A-AZ})] + \mathbf{Tr}(\mathbf{Z}^T\mathbf{Z}) \\
  &=\alpha \mathbf{Tr}(\mathbf{A}^T\mathbf{A}- \mathbf{A}^T\mathbf{A}\mathbf{Z}- \mathbf{Z}^T\mathbf{A}^T\mathbf{A}+ \mathbf{Z}^T\mathbf{A}^T\mathbf{A}\mathbf{Z})+\mathbf{Tr}(\mathbf{Z}^T\mathbf{Z}),
\end{split}
\end{equation}
with its partial derivative being
\begin{equation}
\frac{\partial \mathbf{E}}{\partial \mathbf{Z}} = \alpha(-2\mathbf{A}^T\mathbf{A}+2\mathbf{A}^T\mathbf{A}\mathbf{Z})+2\mathbf{Z}.
\end{equation}
Setting $\partial \mathbf{E}/\partial \mathbf{Z}=0$, we can obtain the optimal solution of $\mathbf{Z}$ as
\begin{equation}
\mathbf{Z}^* = \alpha(\alpha\mathbf{A}^T\mathbf{A}+\mathbf{I})^{-1}\mathbf{A}^T\mathbf{A},
\label{eq_solution_Z}
\end{equation}
where $\mathbf{I}$ is the identity matrix. The likelihood matrix $\mathbf{S}$ can be obtained as
\begin{equation}
\mathbf{S} = \mathbf{AZ}^*.
\label{eq_likelihood_S}
\end{equation}
Then, we rank all nonobserved links in a descending order according to their corresponding values in the likelihood matrix $\mathbf{S}$, with the top-$L$ links constituting the predicted results. The complete procedure of the proposed algorithm as well as an example of a small-size simple network are illustrated in Figure \ref{fig:toy}.

\section*{Results}

To test the algorithm¡¯s accuracy, the set of links, $E$, is randomly divided into two parts: (i) a training set $E^T$, which is treated as known information, and (ii) a probe set (i.e., validation subset) $E^P$, which is
used for testing and can be considered as missing links. No information in the probe set is allowed to be used for prediction, that is to say, in the calculation of $\mathbf{S}$, the adjacency matrix $\mathbf{A}$ only contains links in $E^T$. Obviously, $E^T \cup E^P = E$ and $E^T \cap E^P = \emptyset$. The task of a link prediction algorithm is to uncover the links in the probe set based on the information in the training set.

We adopt two standard metrics to quantify the algorithms' accuracy. The first one is called precision \cite{Herlocker2004}, which is defined as the ratio of relevant elements to the number of selected elements. That is to say, if we take the top-$L$ links as predicted links, among which $L_r$ links are right (i.e., there are $L_r$ links in the probe set $E^P$), then the precision equals $L_r/L$. The second one is called the area under the receiver operating characteristic curve (AUC value for short) \cite{Hanley1982}, which can be interpreted as the probability that a randomly chosen link in $E^P$ (i.e., a missing link that indeed exists but is not observed yet) is ranked higher than a randomly chosen link in $U-E$ (i.e., a nonexistent link), where $U$ is the universal set contains all possible links. If all scores are randomly generated from an independent and identical distribution, the AUC value should be about 0.5. Therefore, the degree to which the value exceeds 0.5 indicates how much the algorithm performs better than pure chance.

\begin{figure}
    \centering
	\includegraphics[width=0.95\textwidth]{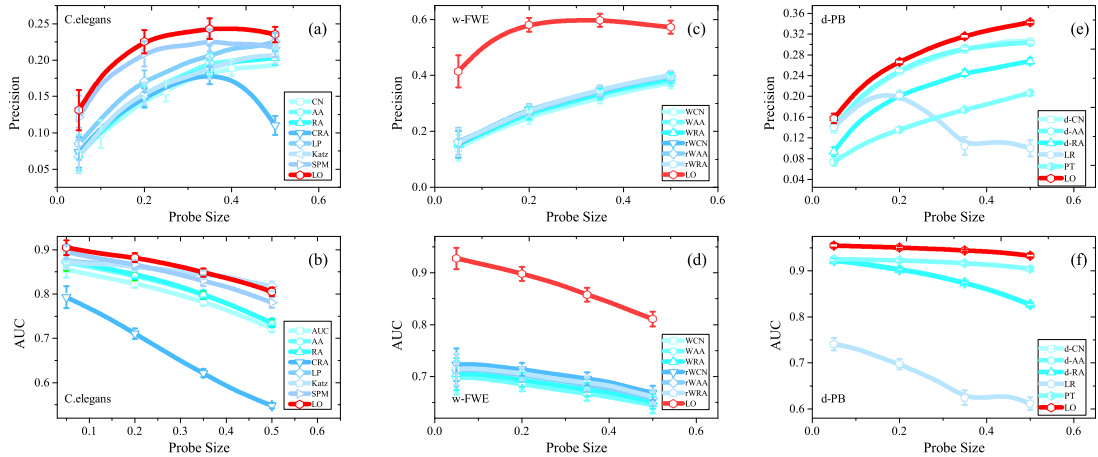}
    \caption{Precision and AUC of the proposed algorithm and the corresponding benchmarks on three selected networks with varying sizes of the probe sets. The results are obtained by 100 independent runs, and the short vertical lines represent standard deviations.}
\label{fig:fig2}
\end{figure}

\subsection*{Simple Networks}

We first test the present linear optimization (LO) method on eight simple networks (i.e., undirected and unweighted networks) from disparate fields, including a food web (FWF), a neural network (C.elegans), a friendship network (Hamster), an air transportation network (USAir), a rating network on movies (MovieRate), a protein-protein interaction network (Reactome), a software dependency network (JDK) and a rating network on Wikepedia (WikiRate). Detailed descriptions and fundamental statistics of these networks are shown in Supplementary Note 2. We compare the proposed method with seven benchmarks, namely the common neighbor (CN) index \cite{Liben-Nowell2007}, the Adamic-Adar (AA) index \cite{Adamic2003}, the resource allocation (RA) index \cite{Ou2007}, the Cannistraci resource allocation (CRA) index \cite{Cannistraci2013}, the local path (LP) index \cite{Lu2009}, the Katz index \cite{Katz1953} and the structural perturbation method (SPM) \cite{Lu2015}. Mathematical details for all benchmark algorithms, including those for weighted networks and directed networks, are presented in Methods.

Table \ref{table_pre_result} compares the prediction accuracy, quantified by precision and AUC, of the proposed algorithm and the seven benchmark algorithms. Obviously, in most cases, LO performs best, usually with remarkably higher accuracy than widely applied local methods (CN, AA, RA, CRA and LP), as well as the famous global index, the Katz index. LO is also slightly better than the state-of-the-art method SPM, and LO runs much faster than SPM (see Supplementary Note 3). We further test the robustness of algorithms' performances by varying the size of probe set from 5\% to 50\%. Again, LO performs overall best. Figures \ref{fig:fig2}(a) and \ref{fig:fig2}(b) show typical results for C.elegans, and all results for the eight simple networks are presented in Supplementary Note 4.

\begin{table}
\footnotesize{
\caption{Precision (top half) and AUC (bottom half) of the proposed algorithm and the seven benchmarks on the eight simple networks. Each result is averaged over 100 independent runs with probe set containing 10\% random links. Parameters in LP, Katz and LO are tuned to their optimal values subject to maximal precision or AUC. The best-performed results are emphasized in bold.}
% title of Table
\centering
\begin{tabular}
{p{2.3cm} p{1.5cm} p{1.5cm} p{1.2cm} p{1.2cm} p{1.2cm} p{1.2cm} p{1.2cm} p{1.2cm}}

% centered columns (4 columns)
\hline        \\[-5.5ex]                  %inserts double horizontal lines
Networks  & CN  & AA & RA & CRA  & LP & Katz & SPM & LO \\
% inserts table
%heading
\hline \\[-5.5ex]
FWF           & 0.071 & 0.073 & 0.074 & 0.076 & 0.298 & 0.152 & 0.574 & \textbf{0.581} \\ 	
C.elegans     & 0.098 & 0.106 & 0.101 & 0.107 & 0.120 & 0.103 & 0.170 & \textbf{0.180}  \\ 		
Hamster       & 0.060 & 0.059 & 0.054 & 0.059 & 0.173 & 0.107 & 0.459 & \textbf{0.485}\\ 		
USAir         & 0.369 & 0.390 & \textbf{0.454} & 0.402 & 0.365 & 0.365 & 0.442 & 0.443 \\ 	
MovieRate     & 0.143 & 0.142 & 0.129 & 0.147 & 0.184 & 0.152 & {0.293} & \textbf{0.297} \\
Reactome      & 0.244 & 0.253 & 0.400 & 0.304 & 0.435 & 0.282 & 0.893 & \textbf{0.907} \\
JDK           & 0.020 & 0.026 & 0.095 & 0.031 & 0.421 & 0.357 & \textbf{0.635} & 0.617 \\
WikiRate      & 0.100 & 0.098 & 0.102 & 0.105 & 0.107 & 0.101 & 0.186 & \textbf{0.193} \\
\hline
FWF           & 0.607 & 0.608 & 0.611 & 0.620 & 0.813 & 0.718 & 0.948 & \textbf{0.950} \\ 	
C.elegans     & 0.854 & 0.862 & 0.866 & 0.768 & 0.853 & 0.851 & 0.884 & \textbf{0.896}  \\ 		
Hamster       & 0.779 & 0.782 & 0.782 & 0.714 & 0.864 & 0.825 & {0.912} & \textbf{0.916} \\ 		
USAir         & 0.936 & 0.947 & \textbf{0.953} & 0.916 & 0.921 & 0.923 & 0.924 & 0.938 \\ 	
Movierate     & 0.902 & 0.904 & 0.902 & 0.903 & 0.920 & 0.910 & 0.945 & \textbf{0.950} \\
Reactome      & 0.989 & 0.990 & 0.991 & 0.978 & 0.992 & 0.991 & 0.991 & \textbf{0.995} \\
JDK           & 0.823 & 0.896 & 0.914 & 0.864 & 0.941 & 0.916 & 0.935 & \textbf{0.983} \\
WikiRate      & 0.927 & 0.928 & 0.928 & 0.882 & 0.954 & 0.947 & 0.941 & \textbf{0.965} \\
\hline %inserts single line
\end{tabular}
\label{table_pre_result}
% is used  to refer this table in the text
}
\end{table}

\begin{table}
\small{
\caption{Precision (top half) and AUC (bottom half) of the proposed algorithm and the six benchmarks on the eight weighted networks. Each result is averaged over 100 independent runs with probe set containing 10\% random links. Parameters in LO are tuned to their optimal values subject to maximal precision or AUC. The best-performed results are emphasized in bold.}
% title of Table
\centering
\begin{tabular}{p{2.5cm} p{1.5cm} p{1.5cm} p{1.5cm} p{1.5cm} p{1.5cm} p{1.5cm} p{1.5cm}}

% centered columns (4 columns)
\hline     \\[-5.5ex]                 %inserts double horizontal lines
Networks  & WCN & \ WAA  & \ WRA & \ rWCN & \ rWAA & \ rWRA & \ LO \\ [0.5ex]
% inserts table
%heading
\hline \\[-5.5ex]

w-FWF         & 0.070 & 0.095 & 0.101 & 0.135 & 0.142 & 0.143 & \textbf{0.578} \\
w-FWE         & 0.179 & 0.177 & 0.181 & 0.200 & 0.193 & 0.196 & \textbf{0.533} \\
w-FWM         & 0.130 & 0.126 & 0.128 & 0.142 & 0.132 & 0.136 & \textbf{0.522} \\
w-C.elegans   & 0.099 & 0.102 & 0.107 & 0.107 & 0.107 & 0.108 & \textbf{0.186} \\
w-USAir       & 0.368 & 0.378 & 0.406 & 0.444 & 0.424 & \textbf{0.447} & 0.444 \\
w-WTN         & 0.422 & 0.425 & 0.447 & 0.454 & 0.451 & 0.450 & \textbf{0.475} \\
w-Macaca      & 0.540 & 0.535 & 0.525 & 0.522 & 0.527 & 0.516 & \textbf{0.739} \\
w-Football    & 0.109 & 0.113 & 0.110 & 0.103 & 0.106 & 0.107 & \textbf{0.241} \\  [0.8ex]
\hline \\[-5.5ex]
w-FWF         & 0.700 & 0.694 & 0.702 & 0.721 & 0.710 & 0.712 & \textbf{0.924} \\
w-FWE         & 0.696 & 0.691 & 0.697 & 0.715 & 0.705 & 0.707 & \textbf{0.927} \\
w-FWM         & 0.710 & 0.706 & 0.711 & 0.716 & 0.711 & 0.714 & \textbf{0.926} \\
w-C.elegans   & 0.827 & 0.830 & 0.834 & 0.833 & 0.834 & 0.833 & \textbf{0.842} \\
w-USAir       & 0.915 & 0.920 & 0.925 & 0.928 & 0.926 & 0.927 & \textbf{0.930} \\
w-WTN         & 0.902 & 0.907 & 0.917 & 0.927 & 0.923 & 0.924 & \textbf{0.931} \\
w-Macaca      & 0.944 & 0.946 & 0.947 & 0.947 & 0.947 & 0.947 & \textbf{0.983} \\
w-Football    & 0.665 & 0.666 & 0.666 & 0.663 & 0.664 & 0.662 & \textbf{0.788} \\  [0.8ex]
\hline %inserts single line
\end{tabular}
\label{table_LOpre_uw}
% is used  to refer this table in the text
}
\end{table}

\subsection*{Weighted Networks}

Many real systems are naturally represented by weighted networks, since the strengths of links are highly heterogeneous and thus the binary representation will lose much information \cite{Newman2004,Barrat2004}. Accordingly, a number of methods are recently proposed to predict missing links in weighted networks \cite{Murata2007,Lu2010,Sa2011,Zhao2015,Sett2016,Zhu2016}. LO can be directly extended to weighted networks via replacing the adjacency matrix   $\mathbf{A}$ by the weight matrix $\mathbf{W}$, where $w_{ij}$ denotes link weight between nodes $i$ and $j$ and $w_{ij}=0$ if $i$ and $j$ are disconnected. To avoid the over contributions from some very strong links, we normalize weights by using a simple sigmoid function as
\begin{equation}
w'=\frac{1}{1+e^{-w}}.
\end{equation}
If all original weights are positive, the normalized weights $w'$ lie in the range $(1/2, 1)$, while in a more general case with negative links \cite{Leskovec2010}, $w'$ lie in the range $(0, 1)$.

\begin{table}
\footnotesize{
\centering
\caption{Precision (top half) and AUC (bottom half) of the proposed algorithm and the five benchmarks on the eight direct networks. Each result is averaged over 100 independent runs with probe set containing 10\% random links. Parameters in LR and LO are tuned to their optimal values subject to maximal Precision or AUC. Due to the relatively high computational complexity of LR, the prediction results for d-WikiRate cannot be obtained in reasonable time. The best-performed results are emphasized in bold.}
\begin{tabular}{p{2.5cm} p{1.8cm} p{1.8cm} p{1.8cm} p{1.8cm} p{1.8cm} p{1.8cm} p{1.8cm} p{1.5cm}}
% centered columns (4 columns)
\hline \\[-5.5ex]             %inserts double horizontal lines
Networks \ & d-CN \ & d-AA \ & d-RA \ & LR \ & PT \ & LO \ \\ [0.5ex]
% inserts table
%heading
\hline \\[-5.5ex]
d-FWF             & 0.056 & 0.075 & 0.088 & 0.546 & 0.111 & \textbf{0.572}  \\ % & 0.5724
d-FWE             & 0.145 & 0.167 & 0.260 & 0.584 & 0.198 & \textbf{0.611}  \\ % & 0.6146
d-FWM             & 0.088 & 0.110 & 0.124 & 0.504 & 0.225 & \textbf{0.526}  \\ % & 0.5115
d-C.elegans       & 0.062 & 0.064 & 0.057 & 0.108 & 0.066 & \textbf{0.137}  \\ % & 0.1419
d-PB              & 0.189 & 0.192 & 0.149 & 0.184 & 0.101 & \textbf{0.227}  \\ % & 0.2052
d-WikiRate        & 0.121 & 0.122 & 0.057 & N/A & 0.073 & \textbf{0.189}  \\ % & 0.1872
d-SmaGrid         & 0.081 & 0.069 & 0.043 & 0.024 & 0.073 & \textbf{0.106}  \\ % & 0.1316
d-SciMet          & 0.054 & 0.044 & 0.023 & 0.035 & 0.049 & \textbf{0.110}  \\ [0.5ex] % & 0.0930
\hline \\ [-5.5ex]
d-FWF             & 0.641 & 0.651 & 0.666 & 0.898 & 0.853 & \textbf{0.969}  \\ % & 0.5724
d-FWE             & 0.750 & 0.758 & 0.762 & 0.917 & 0.810 & \textbf{0.963}  \\ % & 0.6146
d-FWM             & 0.727 & 0.732 & 0.739 & 0.890 & 0.868 & \textbf{0.960}  \\ % & 0.5115
d-C.elegans       & 0.779 & 0.786 & 0.790 & 0.570 & 0.812 & \textbf{0.886}  \\ % & 0.1419
d-PB              & 0.899 & 0.900 & 0.901 & 0.732 & 0.923 & \textbf{0.955}  \\ % & 0.2052
d-WikiRate        & 0.923 & 0.924 & 0.923 & N/A & 0.964 &\textbf{0.973}  \\ % & 0.1872
d-SmaGrid         & 0.702 & 0.699 & 0.705 & 0.545 & 0.818 &\textbf{0.883}  \\ % & 0.1316
d-SciMet          & 0.647 & 0.641 & 0.646 & 0.632 & \textbf{0.861} & 0.797 \\ [0.5ex] % & 0.0930
\hline %inserts single line
\end{tabular}
\label{table:directed}
}
\end{table}

We test the weighted LO method on eight weighted networks (using ''w-" in their names to emphasize), including three food webs (w-FWF, w-FWE and w-FWM), the weighted versions of C.elegant and USAir (w-C.elegant and w-USAir), a world trade network (w-WTN), a cortical neural network (w-Macaca) and a network of football games (w-Football). Detailed descriptions and fundamental statistics of these networks are shown in Supplementary Note 2. We compare the weighted LO with six benchmarks, namely the weighted common neighbor (WCN) index \cite{Murata2007}, the weighted Adamic-Adar (WAA) index \cite{Murata2007}, the weighted resource allocation (WRA) index \cite{Murata2007}, the reliable-route weighted CN (rWCN) index \cite{Zhao2015}, the reliable-route weighted AA (rWAA) index \cite{Zhao2015} and the reliable-route weighted RA (rWRA) index \cite{Zhao2015}. Mathematical definitions are shown in Methods. As shown in Table 2, LO performs best, with remarkably higher accuracy than all other methods. We also test the algorithms' robustness by varying the size of probe set from 5\% to 50\%. Figures 2(c) and 2(d) show typical results for w-FWE and all results for the eight weighted networks are presented in Supplementary Note 4. Again, LO performs best no matter how large the probe size is.

\subsection*{Directed Networks}

Predicting links in directed network is the most challenging problem in link prediction since both the link existence and link direction have to be determined by the algorithm \cite{GTZ2013}. Obviously, LO can be directly extended to directed networks by introducing an asymmetric adjacency matrix $\mathbf{A}$. Recently, a number of methods are proposed to solve this challenge \cite{GTZ2013,ZLW2013,WZZ2015,Xu2017 }. We compare the performance of LO with three kinds of algorithms: (i) the extension of local indices from simple networks to directed networks \cite{Xu2017}, including directed CN (d-CN), directed AA (d-AA) and directed RA (d-RA); (ii) the potential theory (PT) that makes use of local organization principle to predict the existence of missing directed links \cite{ZLW2013}; and (iii) the low rank (LR) approximation algorithm for directed networks \cite{Pech2017}. Mathematical definitions are shown in Methods.

We test the directed LO method as well as the above benchmarks on eight directed networks (using "d-" in their names to emphasize), including directed versions of food webs (d-FWF, d-FWE and d-FWM), C.elegans (d-C.elegans) and WikiRate (d-WikiRate), a network of political blogs (d-PB), and two citation networks (d-SmaGrid and d-SciMet). Detailed descriptions and fundamental statistics of these networks are shown in Supplementary Note 2. As shown in Table 3, LO performs best, with remarkably higher accuracy than all extended indices for directed networks and considerably higher accuracy than PT and LR. We also test the algorithms' robustness by varying the size of probe set from 5\% to 50\%. Figures 2(e) and 2(f) show typical results for d-PB and all results for the eight directed networks are presented in Supplementary Note 4. Again, LO performs best no matter how large the probe size is.

\begin{table}
\footnotesize{
\caption{Precision (top half) and AUC (bottom half) of CN, DLO1 and DLO2 on the eight simple networks. Each result is averaged over 100 independent runs with probe set containing 10\% random links. Parameter in DLO2 is tuned to its optimal values subject to maximal precision or AUC.}
% title of Table
\centering
\begin{tabular}
{p{1.7cm} p{1cm} p{1.5cm} p{1.2cm}<{\centering} p{1.2cm}<{\centering} p{1.5cm} p{1.5cm}<{\centering} p{1.4cm}<{\centering} p{1.2cm}}

% centered columns (4 columns)
\hline        \\[-5.5ex]                  %inserts double horizontal lines
Networks  & FWF  & C.elegans & Hamster  & USAir & MovieRate & Reactome  & JDK & WikiRate \\
% inserts table
%heading
\hline \\[-5.5ex]
CN       & 0.072 &  0.097 & 0.060 & 0.369 & 0.143 & 0.244 & 0.024 & 0.100 \\ 	
DLO1     & 0.315 &  0.123 & 0.185 & 0.358 & 0.188 & 0.445 & 0.511 & 0.107 \\ 		
DLO2     & 0.474 & 0.160 & 0.363 & 0.442 & 0.233 & 0.498 & 0.516 & 0.116 \\ 		
\hline
CN       & 0.605 & 0.851 & 0.778 & 0.936 & 0.905 & 0.989 & 0.823 & 0.927 \\ 	
DLO1     & 0.816 & 0.846 & 0.860 & 0.897 & 0.923 & 0.987 & 0.947 & 0.954 \\ 		
DLO2     & 0.921 & 0.892 & 0.903 & 0.939 & 0.935 & 0.991 & 0.955 & 0.961 \\ 		
\hline %inserts single line
\end{tabular}
\label{table_DLO_result}
% is used  to refer this table in the text
}
\end{table}

\begin{table}
\footnotesize{
\caption{$P^{(2)}$, $P^{(3)}$, $S^{(2)}$ and $S^{(3)}$ for the eight simple networks.}
% title of Table
\centering
\begin{tabular}
{p{1.7cm} p{1cm}<{\centering} p{1.4cm}<{\centering} p{1.4cm}<{\centering} p{1.4cm}<{\centering} p{1.5cm} p{1.4cm}<{\centering} p{1.4cm}<{\centering} p{1.2cm}}

% centered columns (4 columns)
\hline        \\[-5.5ex]                  %inserts double horizontal lines
Networks  & FWF   & C.elegans & Hamster  & USAir & MovieRate & Reactome  & JDK & WikiRate \\
% inserts table
%heading
\hline \\[-5.5ex]
$P^{(2)}$                 & 1       & 0.996 & 0.9888 & 0.9954 & 0.999	& 0.9972 & 1 & 0.9933\\ 	
$P^{(3)}$                 & 0.9668  & 0.5755 & 0.5367 & 0.4739 & 0.7796 & 0.2560 & 0.8858 & 0.2182 \\
$S^{(2)}$         & 0.9552  & 0.6771 & 0.6871 & 0.6008 & 0.9048 & 0.1493 & 0.5473 & 0.25 \\ 	
$S^{(3)}$         & 0.9987  & 0.971 & 0.975 & 0.9496 & 0.9989 & 0.5135 & 0.9836 & 0.8395 \\		
\hline %inserts single line
\end{tabular}
% is used  to refer this table in the text
}
\end{table}

\subsection*{Degenerated Local Indices}

After observing the remarkably higher prediction accuracy of LO than other well-known local indices, we would like to uncover the underlying mechanism resulting in LO's advantage. By substituting $\mathbf{Z}^*$ in Eq. (7), one obtains
\begin{equation}
\begin{split}
    \mathbf{S} & = \mathbf{AZ}^* \\
    & = \mathbf{A}(\mathbf{A}^T\mathbf{A}+\frac{1}{\alpha}\mathbf{I})^{-1}\mathbf{A}^T\mathbf{A} \\
     & = \mathbf{A}(\mathbf{A}^T\mathbf{A}+\frac{1}{\alpha}\mathbf{I})^{-1}(\mathbf{A}^T\mathbf{A}+\frac{1}{\alpha}\mathbf{I}-\frac{1}{\alpha}\mathbf{I}) \\
     & = \mathbf{A}-\mathbf{A}(\mathbf{I}+\alpha \mathbf{A}^T\mathbf{A})^{-1}.
\end{split}
\end{equation}
Applying the Neumann series, if $\alpha < 1/\lambda_{max}^2$ ($\lambda_{max}$ is the largest eigenvalue of matrix A),  
\begin{equation}
(\mathbf{I}+\alpha \mathbf{A}^T\mathbf{A})^{-1}=\mathbf{I} - \alpha\mathbf{A}^2 + \alpha^2\mathbf{A}^4 - \alpha^3\mathbf{A}^6 + \cdots, 
\end{equation}
and thus Eq. (9) can be rewritten as
\begin{equation}
\begin{split}
    \mathbf{S}
               & = \mathbf{A} - \mathbf{A}(\mathbf{I} - \alpha\mathbf{A}^2 + \alpha^2\mathbf{A}^4 - \alpha^3\mathbf{A}^6 + \cdots) \\
               & = \alpha\mathbf{A}^3 - \alpha^2\mathbf{A}^5 + \alpha^3\mathbf{A}^7 - \alpha^4\mathbf{A}^9 + \cdots.
\end{split}
\end{equation}
Comparing with the famous Katz index (i.e., $\beta\mathbf{A} + \beta^2\mathbf{A}^2+ \beta^3\mathbf{A}^3 +\cdots$, where the first term doesn't work since all unobserved links correspond to zero elements in $\mathbf{A}$), the differences lie in three aspects: (i) The expansion of LO starts from $\mathbf{A}^3$ (i.e., the number of 3-hop paths), instead of the usually considered item $\mathbf{A}^2$ (i.e., the number of 2-hop paths or the number of common neighbors); (ii) LO only takes into account odd paths; (iii) Some items in Eq. (10) play negative roles.
To look closer, we focus on two degenerated local indices from LO, say $\mathbf{A}^3$ (named as DLO1) and $\mathbf{A}^3-\alpha \mathbf{A}^5$ (named as DLO2). Table 4 compares the prediction accuracy of CN, DLO1 and DLO2 on the eight simple networks. Two observations are highly striking.

First of all, DLO1 remarkably outperforms CN, which challenges our intuition that shorter paths indicate stronger correlation than longer paths \cite{Katz1953,Lu2009}. This is largely due to two following reasons. Firstly, DLO1 (i.e., the number of 3-hop paths) is more informative than CN (i.e., the number of 2-hop paths). Denoting $e^{(2)}$ the set of node pairs connected by at least one 2-hop path and $e^{(3)}$ the set of node pairs connected by at least one 3-hop path, then we calculate the fraction of node pairs connected by 3-hop paths in the set of node pairs having common neighbors $P^{(2)} = |e^{(2)}\bigcap e^{(3)}|/|e^{(2)}|$, as well as the fraction of node pairs connected by 2-hop paths in the set of node pairs connected by 3-hop paths $P^{(3)} = |e^{(2)}\bigcap e^{(3)}|/|e^{(3)}|$. As shown in Table 5, for all the eight networks, $P^{(3)}<P^{(2)}$ with $P^{(2)}$ very close to 1. That is to say, almost all node pairs being connected by at least one 2-hop path are also connected by at least one 3-hop path, while a considerable portion of node pairs connected by 3-hop paths do not have common neighbors. Hence we say $A^3$ is more informative than $A^2$. Secondly, DLO1 is more distinguishable than CN. Denoting $N$ the number of nodes in the target network and $n^{(l)}_i$ the number of node pairs connected by $i$ different $l$-hop paths (these paths are allowed to pass through a node multiple times), then the ratio of node pairs connected by $i$ different $l$-hop paths is $R^{(l)}_i=2n^{(l)}_i / N(N-1)$ as there are in total $N(N-1)/2$ node pairs. L\"u \emph{et al.} \cite{Lu2009} showed an extremal case that in the router-level Internet \cite{Spring2004} 99.59\% of node pairs do not have common neighbors and 91.11\% of those having common neighbors have just 1 common neighbor. In such case, the CN index is not distinguishable and the corresponding distribution of $R^{(2)}$ is highly concentrated. Therefore, we apply the famous diversity measure, called Simpson coefficient \cite{Simpson1949}, to quantify the distinguishabilities of DLO1 and CN, namely
\begin{equation}
S^{(l)}=1-\sum_i [R^{(l)}_i]^2,
\end{equation}
where $i$ runs from zero to its possibly maximum value. Obviously, the larger $S$ corresponds to more diverse and thus more distinguishable distribution of $R$. As shown in Table 5, $A^3$ is more distinguishable than $A^2$ (direct comparison between distributions of $R^{(2)}$ and $R^{(3)}$ is presented in Supplementary Note 5). Putting the above two reasons together, it is now not surprising that the number of 3-hop paths is a better index than common neighbors in link prediction.

Secondly, DLO2 remarkably outperforms DLO1. Clearly, as suggested by the well-known Homophily mechanism \cite{McPherson2001}, if two nodes share many features, they have high probability to be directly connected \cite{Kossinets2006}. Notice that, two nodes are probably connected by many 3-hop paths, and these paths may be built from independent reasons or may contain redundant information. The former usually indicates a higher similarity between the two nodes and thus to eliminate redundant correlation can improve the accuracy of link prediction \cite{ZSL2009}. Figure 3 shows two examples where nodes $i$ and $j$ are both connected by 4 3-hop paths, say $(\mathbf{A}^3)_{ij}=4$. The 4 paths in Figure 3(a) are independent while the 4 paths in Figure 3(b) are overlapped. Indeed, in the latter case, there are only two independent paths connecting $i$ and $j$. At the same time, the overlapping paths will result in densely connected local structure and thus larger value of $(\mathbf{A}^5)_{ij}$, since $(\mathbf{A}^5)$ includes the paths passing through a node by multiple times. Therefore, larger value of $(\mathbf{A}^5)_{ij}$ indicates denser local connections and thus more redundance. This is the reason why to punish node pairs with many 5-hop paths will lead to better prediction as DLO2.

In a word, we strongly suggest DLO1 ($\mathbf{A}^3$) and DLO2 ($\mathbf{A}^3-\alpha \mathbf{A}^5$) two very good quasi-local indices for link prediction.

\begin{figure}
    \centering
	\includegraphics[width=0.95\textwidth]{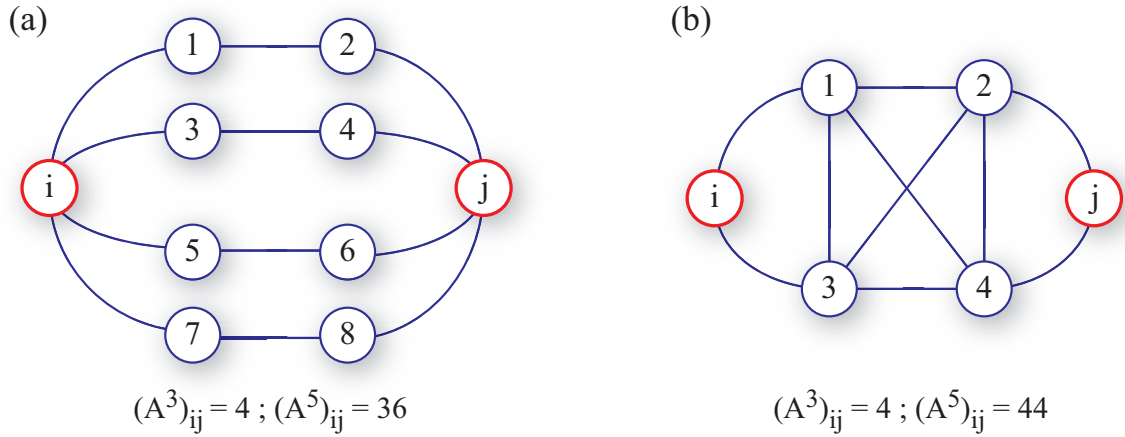}
    \caption{The illustration of redundant information.}
\label{fig:fig5}
\end{figure}

\section*{Discussion}

This work starts from a very simple assumption that the likelihood of the existence of a link between two nodes can be unfolded by a linear summation of contributions of their common neighbors. The optimal likelihood matrix can be analytically obtained, with remarkably higher prediction accuracy than other state-of-the-art algorithms. The solution can be directly extended to weighted and directed networks, also with much better performance than well-known benchmarks. In particular, link prediction in directed networks is a great challenge and the proposed LO algorithm shows a huge advantage as shown in Table 3.

It is very interesting to notice that a formula similar to Eq. (6), named as ridge regression \cite{Hoerl1970,Tikhonov1977}, was long ago proposed to estimate the solution of $X$ in linear equations $Y=\mathbf{A}X+\varepsilon$ with $\mathbf{A}$ a singular matrix and $\varepsilon$ the noise. Though the details of solutions of the two problems are different (e.g., the present solution does not involve noise $\varepsilon$ or vector $Y$), both method consider the usage of norm regularization to avoid the overfitting. Such regularization has recently found significant applications in disparate fields, such as brain science \cite{Smith2015} and artificial intelligence \cite{Silver2017}. Hence we believe the present linear optimization method could also find wide applications in graph mining and matrix completion.

Lastly, after finishing this work, we are happy to see strongly supportive experiments in a very recent preprint \cite{BioRxiv2018}, which shows that the number of 3-hop paths (named as $L_3$, similar to DLO1) significantly outperforms CN index in predicting protein-protein interactions across multiple real datasets. Beyond biological explanations \cite{BioRxiv2018}, our work indeed provides a solid theoretical basis with a more universal perspective. In the future work, we intend to compare the degenerated local indices with other local methods based on extensive real data.

\section*{Methods}

This section presents three categories of benchmark algorithms. The first category is for simple networks, including the common neighbor (CN) index \cite{Liben-Nowell2007}, the Adamic-Adar (AA) index \cite{Adamic2003}, the resource allocation (RA) index \cite{Ou2007}, the Cannistraci resource allocation (CRA) index \cite{Cannistraci2013}, the local path (LP) index \cite{Lu2009}, the Katz index \cite{Katz1953} and the structural perturbation method (SPM) \cite{Lu2015}. CN index is defined as
\begin{equation} \label{eq_cn}
    S_{xy}^{CN}=|\Gamma(x) \cap \Gamma(y)|,
\end{equation}
where $\Gamma(x)$  and $\Gamma(y)$ are sets of neighbors of nodes $x$ and $y$, respectively. AA and RA indices assign small-degree neighbors more weights, as
\begin{equation} \label{eq_aa}
    S_{xy}^{AA}=\sum_{s\in|\Gamma(x) \cap \Gamma(y)|}\frac{1}{\log(|\Gamma(s)|)},
\end{equation}
and
\begin{equation} \label{eq_ra}
    S_{xy}^{RA}=\sum_{s\in|\Gamma(x) \cap \Gamma(y)|}\frac{1}{|\Gamma(s)|}.
\end{equation}
LP index considers both contributions from 2-hop and 3-hop paths, as
\begin{equation}
    S_{xy}^{LP} = (\mathbf{A}^2)_{xy} + \epsilon(\mathbf{A}^3)_{xy},
\label{eq_LP}
\end{equation}
where $\epsilon$ is a free parameter. Katz index considers all possible paths connecting nodes $x$ and $y$ with exponentially damped weights, as
\begin{equation} \label{eq_katz}
    S_{xy}^{Katz}=\beta(\mathbf{A})_{xy}+\beta^2(\mathbf{A}^2)_{xy}+\beta^3(\mathbf{A}^3)_{xy}+\cdots,
\end{equation}
where $\beta$ is a free parameter. The SPM splits the observed network into two parts: a background network containing most links and a perturbation network containing a small portion of links. It uses eigenvectors of the background network while eigenvalues of the observed network to approximately reconstruct the observed network and the large-value elements in the reconstructed network but not the observed network indicate missing links. Readers are encouraged to find the mathematical details in the original article \cite{Lu2015}.

The second category is for weighted networks, including the weighted common-neighborhood-based indices \cite{Murata2007} (i.e., WCN, WAA and WRA) and the reliable-route weighted indices \cite{Zhao2015} (i.e., rWCN, rWAA and rWRA). They are mathematically defined as follows.
\begin{equation}
    \mathbf{S}_{xy}^{WCN}=\sum_{s\in|\Gamma(x) \cap \Gamma(y)|}(w_{xs}+w_{sy}),
\label{eq_WCN}
\end{equation}
\begin{equation}
    \mathbf{S}_{xy}^{WAA}=\sum_{s\in|\Gamma(x) \cap \Gamma(y)|}\frac{w_{xs}+w_{sy}}{\log(1+|\Gamma(s)|)},
\label{eq_WAA}
\end{equation}
\begin{equation}
    \mathbf{S}_{xy}^{WRA}=\sum_{s\in|\Gamma(x) \cap \Gamma(y)|}\frac{w_{xs}+w_{sy}}{|\Gamma(s)|},
\label{eq_WRA}
\end{equation}
\begin{equation}
    \mathbf{S}_{xy}^{rWCN}=\sum_{s\in|\Gamma(x) \cap \Gamma(y)|}(w_{xs} \cdot w_{sy}),
\label{eq_rWCN}
\end{equation}
\begin{equation}
    \mathbf{S}_{xy}^{rWAA}=\sum_{s\in|\Gamma(x) \cap \Gamma(y)|}\frac{w_{xs}\cdot w_{sy}}{\log(1+|\Gamma(s)|)},
\label{eq_rWAA}
\end{equation}
\begin{equation}
    \mathbf{S}_{xy}^{rWRA}=\sum_{s\in|\Gamma(x) \cap \Gamma(y)|}\frac{w_{xs}\cdot w_{sy}}{|\Gamma(s)|}.
\label{eq_rWRA}
\end{equation}

The third category is for directed networks, including the directed common-neighborhood-based indices \cite{Xu2017} (i.e., d-CN, d-AA, d-RA), the low rank matrix completion method (LR method for short) for directed networks \cite{Pech2017}, and the potential theory (PT) \cite{ZLW2013}. The directed common-neighborhood-based indices are defined as
\begin{equation}
    S_{xy}^{d-CN}=|\Gamma^{out}(x) \cap \Gamma^{in}(y)|,
\end{equation}
\begin{equation}
    S_{xy}^{d-AA}=\sum_{s\in|\Gamma^{out}(x) \cap \Gamma^{in}(y)|}\frac{1}{\log(|\Gamma^{out}(s)|)},
\end{equation}
\begin{equation}
    S_{xy}^{d-RA}=\sum_{s\in|\Gamma^{out}(x) \cap \Gamma^{in}(y)|}\frac{1}{|\Gamma^{out}(s)|},
\end{equation}
where $\Gamma^{out}(x)$ is the set of nodes that $x$ points to, $\Gamma^{in}(y)$ is the set of nodes pointing to $y$, and $S_{xy}$ here denotes the likelihood of a directed link from $x$ to $y$. The LR method decomposes the adjacency matrix into a low-rank matrix and a sparse matrix, where the former contains missing links and the latter contains spurious links. More details are presented in the original article \cite{Pech2017}. PT assumes that the motifs obeying the potential theory are preferred, and thus links generating more preferred motifs are of higher likelihoods. Mathematical details can be found in the original article \cite{ZLW2013}.

\bibliographystyle{naturemag}

\end{document}